\documentclass[conference]{IEEEtran}
\IEEEoverridecommandlockouts
\usepackage{cite}
\usepackage{amsmath,amssymb,amsfonts}
\usepackage{algorithmic}
\usepackage{graphicx}
\usepackage{textcomp}
\usepackage{xcolor}
\def\BibTeX{{\rm B\kern-.05em{\sc i\kern-.025em b}\kern-.08em
    T\kern-.1667em\lower.7ex\hbox{E}\kern-.125emX}}
    
\def\peter{\textcolor{black}}

\def\giang{\textcolor{black}}
\def\lewis{\textcolor{black}}
\def\caitlin{\textcolor{black}}

\begin{document}

\title{SMERC: Social media event response clustering using textual and temporal information}

\author{\IEEEauthorblockN{Peter Mathews\IEEEauthorrefmark{1},
Caitlin Gray\IEEEauthorrefmark{2}, Lewis Mitchell\IEEEauthorrefmark{3}\IEEEauthorrefmark{4},
Giang T.~Nguyen\IEEEauthorrefmark{5}, and Nigel G.~Bean\IEEEauthorrefmark{6}}
\IEEEauthorblockA{School of Mathematical Sciences,\\
ARC Centre of Excellence for Mathematical \& Statistical Frontiers (ACEMS),\\
The University of Adelaide, SA 5005, Australia\\
\IEEEauthorrefmark{1}peter.mathews@adelaide.edu.au,
\IEEEauthorrefmark{2}caitlin.gray@adelaide.edu.au,\\
\IEEEauthorrefmark{3}Data to Decisions CRC stream lead,
\IEEEauthorrefmark{4}lewis.mitchell@adelaide.edu.au,\\
\IEEEauthorrefmark{5}giang.nguyen@adelaide.edu.au,
\IEEEauthorrefmark{6}nigel.bean@adelaide.edu.au}}

\maketitle

\begin{abstract}
Tweet clustering for event detection is a powerful modern method to automate the real-time detection of events. In this work we present a new tweet clustering approach, using a probabilistic approach to incorporate temporal information. By analysing the distribution of time gaps between tweets we show that the gaps between pairs of related tweets exhibit exponential decay, whereas the gaps between unrelated tweets are approximately uniform. Guided by this insight, we use probabilistic arguments to estimate the likelihood that a pair of tweets are related, and build an improved clustering method. Our method \emph{Social Media Event Response Clustering (SMERC)} creates clusters of tweets based on their tendency to be related to a single event. We evaluate our method at three levels: through traditional event prediction from tweet clustering, by measuring the improvement in quality of clusters created, and also comparing the clustering precision and recall with other methods. By applying SMERC to tweets collected during a number of sporting events, we demonstrate that incorporating temporal information leads to state of the art clustering performance.
\end{abstract}


\lewis{Postings to social media platforms are increasingly being used to extract useful information about real-world events.
Journalists use platforms such as Twitter to} \giang{summarize} \lewis{breaking news stories \cite{Liu:2016:RTL:2983323.2983363},  governments are interested in mining social media to provide early warning of events such as disease outbreaks \cite{StLouise2353}, civil unrest events \cite{vanNoord2017} and even natural disasters such as earthquakes \cite{Sakaki10earthquakeshakes}.
Developing methods to summarize the large volumes of information generated on social media by such events is therefore of great importance for scientists and end-users alike, and an extensive amount of literature has appeared on real-time microblog summarization for event detection.}

\lewis{Sporting events provide a unique testing ground for event detection, due to the collective attention of often large audiences on the in-game action.
While event detection techniques can be applied to any type of event, sporting events are ideal to collect and study as the time, location and hashtags are generally known beforehand.
Furthermore, online reaction to sporting events can be useful case studies on human behaviour, with heavy levels of engagement and emotion generally organized around the success of opposing teams.}

\lewis{In this work we develop a new method for clustering microblog posts authored in response to real-world events.
Distinguishing it from previous methods, our proposed methodology uses both the timing and content of messages to estimate a probability that messages are related,
giving more meaningful event clusters than previous methods.
Through a detailed investigation of the distributions of related and unrelated messages around particular events, we arrive at a probabilistic method for clustering messages.
This approach also provides insight into the mechanisms causing a social media response to a particular event and allows us to model the distribution of response} 
\caitlin{ times.}

\peter{Our work makes the following new contributions:}
\begin{itemize}
	\item Showing that the time gaps between pairs of related messages are exponentially distributed.
	\item Presenting a novel way to improve tweet clustering by incorporating both temporal and textual information.
\end{itemize}

We apply our method \emph{Social Media Event Response Clustering} (SMERC) to three Twitter datasets collected from cricket and Australian Rules Football matches, \lewis{and demonstrate that the clusters around events obtained are meaningful and that our event detection improves over existing techniques.}

\section{Previous work}

\peter{There exists a broad literature on microblog post summarization for both trend and event detection, particularly for sporting events. Tweet clustering} \caitlin{is often focused on using tweet content and user features, but} \peter{has limitations based on the amount of information contained in a tweet.} Studies typically use either \emph{textual} features such as tracking the number of pre-defined keywords or hashtags, or purely \emph{temporal} features such as the timing of posts. \caitlin{There exist only a small number of studies that utilise both textual and temporal features in a meaningful way \cite{doi:10.1177/0165551517698564}.}

\peter{Yang and Leskovec \cite{Yang:2011:PTV:1935826.1935863} summarized two key components for time series clustering} \caitlin{in online media,} \peter{a distance measure and a clustering algorithm. The most commonly used distance measure is Euclidean distance, which has been used in a variety of works, e.g. \cite{DBLP:journals/corr/BertholdH16, gavrilov2000mining}.} \peter{More advanced measures include Dynamic Time Warping \cite{Berndt:1994:UDT:3000850.3000887} and the Longest Common Subsequence \cite{Keogh2005}. The most common clustering algorithm continues to be k-means clustering \cite{Macqueen67somemethods} despite the limitations of having to specify the number of clusters beforehand, and being sensitive to the starting point.}

\peter{Leskovec \emph{et al.} \cite{Leskovec.Backstrom.ea2009Meme-trackingandDynamics} analyzed how popularity of memes varies over time. Our work focuses on a similar problem, but over much shorter time scales. Memes tend to be popular for months while Twitter users tend to respond to events or tweets in seconds or minutes.} This makes our context arguably more challenging.

Most apparently similar to this work, \peter{Gillani \emph{et al.} \cite{Gillani:2017:PSM:3041021.3054146} provided a way to identify key events in sporting contests by clustering both temporal and textual features. They use a threshold technique to determine whether an event is sufficiently significant, and incorporate post time as a feature in a relatively straightforward manner, by appending it to a vector of word counts collected within a window.}  \peter{As this method uses k-means for clustering and it is not possible to know the number of key events beforehand, it is perhaps more suited to the problem of \emph{post-hoc} microblog summarization rather than real-time event detection.}
Conversely, as we show, our approach will be more suited to real-time tweet clustering and event detection;
by reasoning more probabilistically about the distributions of times between tweets around events,
we develop a mathematical model for incorporating temporal information into tweet clustering instead of simply adding it as part of a feature vector.

%\subsection{Tweet clustering and using Twitter for event detection}

\peter{There exist many methods for using Twitter for event detection. Twevent \cite{Li:2012:TSE:2396761.2396785} extracts continuous and non-overlapping word segments, and then calculates bursty event segments within a fixed length window. To evaluate their method, they defined `precision' as the proportion of detected events related to realistic events, and `recall' as the proportion of realistic events detected from the data set. 
Due to the difficulty in assessing whether or not a particular tweet, or cluster of tweets, is related to an event or what constitutes a `realistic' event,
such definitions are simultaneously important and hard to define.
For comparison with previous work we employ the same definitions as used to evaluate Twevent in this study, 
but will also examine the quality of the clusters associated with individual events on a tweet-by-tweet level through studying the content of example tweet clusters.}

\peter{Several authors have used Twitter as a method for social sensing in sporting and other events, e.g. \cite{DBLP:journals/corr/abs-1106-4300, 10.1371/journal.pone.0144646, Gillani:2017:PSM:3041021.3054146}. Zhao \emph{et al.} \cite{DBLP:journals/corr/abs-1106-4300} used the Twitter response to events in NFL games to identify key game events such as touchdowns, interceptions, fumbles and field goals. }\caitlin{The temporal component of} \peter{their event detection method was based on post rate in time windows, using the fact that social media activity increases heavily after key events. They also used content-based keywords to determine whether keyword frequency was above a pre-defined threshold.}

\peter{Tweet intensity is generally modelled as a self exciting temporal point process \cite{Zhao:2015:SSP:2783258.2783401, Zhu:2013:PUA:2505515.2505518}. Events and other tweets tends to drive other activity, leading to burstiness. The rate of twitter activity is affected by other factors such as human prioritization of tasks and circadian cycles \cite{12207979520170327, Mathews2017}}.
%\cite{12207979520170327}

\section{Datasets and analysis}

In this section \caitlin{we introduce the datasets used in the study and perform an analysis of the distributions of temporal information in tweets which form the basis for SMERC.}

\subsection{Datasets}

\peter{We collected tweets around the Australian Big Bash cricket league and Australian Rules Football League (AFL) based on predetermined hashtags for the events, using the Twitter streaming API. The hashtags used and number of tweets collected are outlined below.
As with many sporting events, cricket matches and AFL games have specific hashtags, usually the names of the teams involved, that are publicized by the clubs and supporters prior to the event.
While our method is for tweet clustering generally, such events are ideal for the present study as data collection can be planned in advance.
Our collected data was stored in a MongoDB database and was later processed using custom Python 3.6 scripts.}

\peter{To justify our temporal adjustment, we closely analyze the data from}
\caitlin{three Twitter datasets:}
\peter{
	\begin{itemize}
		\item Dataset D1: Tweets from the Australian Rules Football preliminary final match between the Adelaide Crows and the Geelong Cats on 22 September 2017. \\
		Hashtag: \#AFLCrowsCats \\
		Number of tweets collected: 5018
		\item Dataset D2: Tweets from the 2017/18 Big Bash (the Australian men's domestic Twenty20 cricket tournament) between Brisbane Heat and Melbourne Stars on 20 December 2017. \\
		Hashtag: \#BBL07 \\
		Number of tweets collected: 3153
		\item Dataset D3: Tweets from the 2017/18 Women's Big Bash (the Australian women's domestic Twenty20 cricket tournament) opening weekend on 9/10 December 2017. \\
		Hashtag: \#WBBL03 \\
		Number of tweets collected: 5393
	\end{itemize}
}
%We make our datasets public at https://github.com/pete1729/temporal-clustering. \lm{Might need to withhold this for the submission for the purposes of double-blindage...}
We remark that similar results to the following were obtained across all sporting events studied.
Our datasets are stored in a github repository which will be made public upon acceptance of our paper.

\subsection{Probability of tweets being related exhibits exponential decay with time difference}
\label{sec:ProbabilisticMotivation}

We manually labelled tweets from datasets D1 to D3 in order to examine how the probability of tweets being related varies depending on the time gap between them.
As we will show, this decays exponentially, an observation which will form a critical part of the clustering algorithm SMERC we develop in Section \ref{sec:clustering}.

\peter{Our methodology to analyze the temporal relationship between related tweets is as follows:}
\peter{
	\begin{enumerate}
		\item Collect a series of tweets with a selected hashtag over a period of time.
		\item Identify a set of on-field events (scoring events, penalties, etc), then manually cluster tweets that are in response to these events.
		\item Record the time gaps between pairs of tweets where both tweets are in response to the same manually clustered event.
		\item Record the time gaps between pairs of tweets, where only one of the tweets is in response to a manually clustered event, while the other is unrelated.
		%\item Calculate the text similarity scores between pairs of tweets (as subsequently discussed in Section \ref{sec:clustering}).
		\item Bucketize the data and count the number of related pairs of tweets and unrelated pairs of tweets within each bucket.
		\item Consequently measure the likelihood that a pair of tweets within a bucket is related, and examine how this varies with the time difference.
	\end{enumerate}
}

%After we have the time gaps and text similarity scores between pairs of related tweets and pairs of unrelated tweets, we estimate the corresponding probability densities using a 1D kernel density estimation. We implement this with the KernelDensity package from the sklearn.neighbors module in Python 3.6 \cite{scikit-learn}.

\begin{figure}
	\centering	
	\includegraphics[width=8cm]{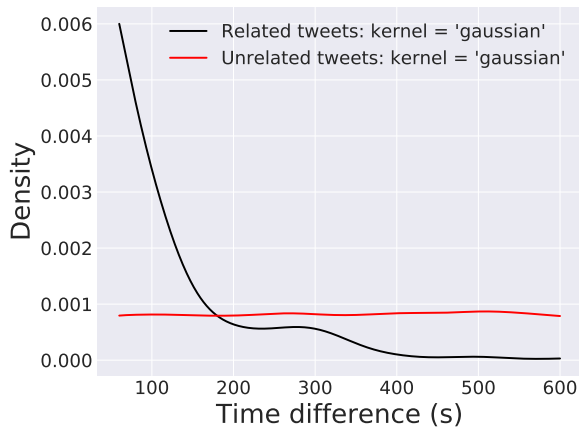}
	\caption{AFL dataset D1: Density of time differences between pairs of related and unrelated tweets. The density of time differences between pairs of unrelated tweets stays roughly constant, while the density of time differences between related tweets decays. The slight bump at around 300 seconds is likely due to noise.}
	\label{fig:AFLRelatedUnrelated}
\end{figure}

\begin{figure}
	\centering	
	\includegraphics[width=8cm]{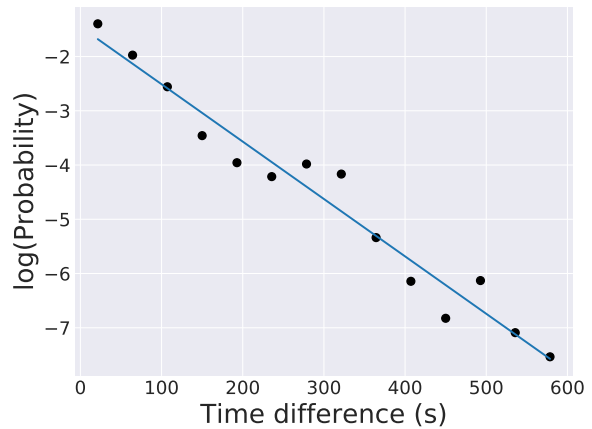}
	\caption{AFL dataset D1: Log-linear plot of the probability of tweets being related, given a time separation. The straight line indicates exponential decay.}
	\label{fig:AFLLogLinear}
\end{figure}

We first analyse dataset D1, the response to events in an AFL game. After manually classifying tweets, we have 18175 pairs of related tweets and 747779 pairs of unrelated tweets. 
We fit a Gaussian kernel to the associated time gaps in order to create a smooth density curve, implemented with the KernelDensity package from the sklearn.neighbors module in Python 3.6 \cite{scikit-learn}.
The resultant density curves of time gaps between pairs of related and unrelated tweets are shown in Figure \ref{fig:AFLRelatedUnrelated}. 
%From tests with the tophat and epanechnikov kernels \cite{citeulike:13887103}, we note that the curve shape is roughly independent of the choice of kernel. 
As can be seen, the density of time gaps between unrelated tweets remains approximately constant with time difference, while the density of gaps between two related tweets about a given event decays over time.
\peter{To determine the probability of tweets being related to the same event given the time separation, we bucketize the data and compute the proportion of related pairs of tweets to overall number of pairs within each bucket. We then plot this on a log-linear plot. As can be seen in Figure \ref{fig:AFLLogLinear}, we have roughly a straight line, corresponding to exponential decay.} The curve has a slope of $-0.01058$, giving that a suitable model for the probability that tweet pairs with gap time $\Delta t$ are related is $Ce^{-0.01058 \Delta t}$, where $C$ is a constant.

\begin{figure}
	\centering	
	\includegraphics[width=8cm]{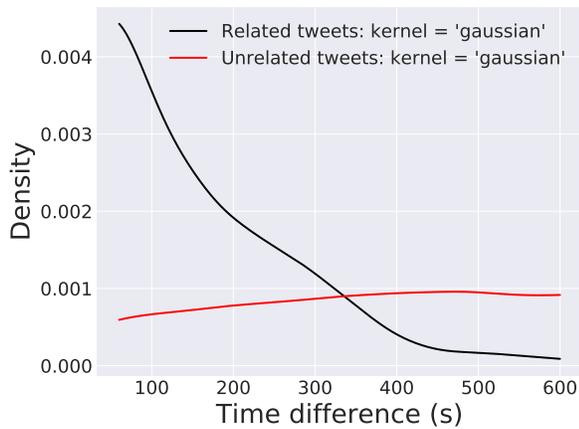}
	\caption{WBBL dataset D3: Density of time differences between pairs of related and unrelated tweets. The density of time differences between pairs of unrelated tweets stays roughly constant with time difference, while the density of time differences between related tweets decays.}
	\label{fig:WBBLRelatedUnrelated}
\end{figure}

\begin{figure}
	\centering	
	\includegraphics[width=8cm]{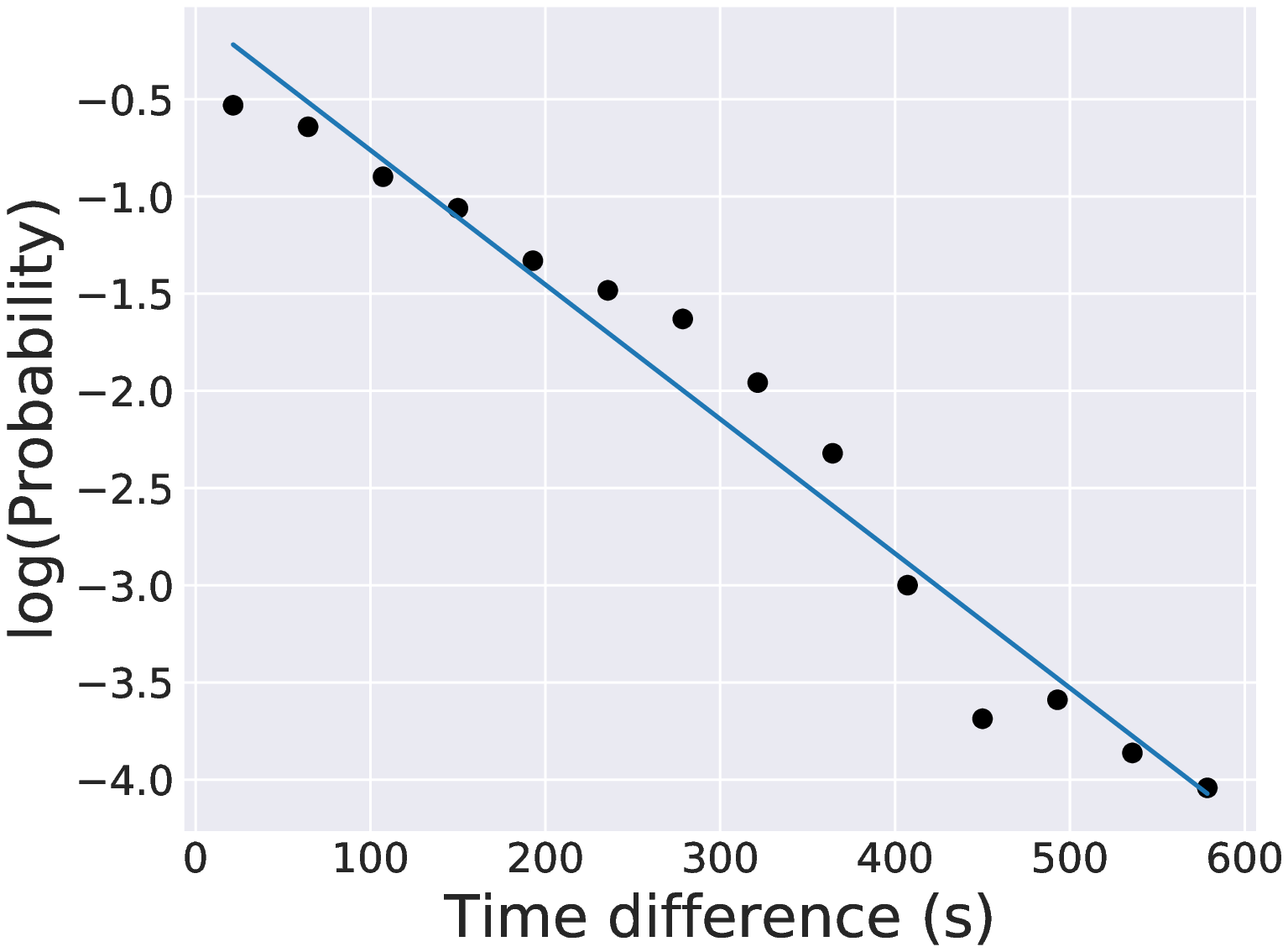}
	\caption{WBBL dataset D3: Log-linear plot of the probability of tweets being related, given a time separation. The straight line indicates exponential decay.}
	\label{fig:WBBLLogLinear}
\end{figure}

In order to demonstrate that this property is not unique to AFL and instead holds across different sports, we repeat here the same analysis for the WBBL dataset D3.
For the WBBL dataset, we have 7430 pairs of related tweets and 50431 pairs of unrelated tweets.
As shown in Figure \ref{fig:WBBLRelatedUnrelated}, we again have that the density of time gaps between unrelated tweets remains approximately constant with time difference, while the density of gaps between two related tweets about a given event decays over time. Bucketizing the data and plotting the probability curve in Figure \ref{fig:WBBLLogLinear} shows that a similar curve as fit to dataset D1 describes the data well. The fitted curve has slope $-0.00692$, slightly less steep than for the AFL dataset D1. This indicates that for both datasets, the change in probability of tweets being related has an exponential relationship with the time gap between them. This is somewhat remarkable, due to the highly different nature of action for the two sports. Similar to baseball, cricket is a fundamentally `discrete' game, whereas like soccer, AFL is fundamentally `continuous'. Note that a similar exponential relationship was also found to hold in dataset D2, as well as for the other sporting events collected (not shown).\\

%\begin{comment}
%\peter{Let A represent the event that two tweets are related. Let B represent if two tweets are separated by time $t$. We have 
%\begin{equation}
%P(A|B) = \frac{ae^{-at}}{c+ae^{-at}}.
%\end{equation}
%%As $t \to \infty, P(A|B) \to (a/c)e^{-at}$, a decaying exponential.}
%\end{comment}

We have shown that the probability of tweets being related decays exponentially with the time gap between them. The exponential decay is likely explained by previous findings that human interest in topics decays exponentially over time \cite{Li:2014:MBT:2566267.2566312}. This motivates a key new step in our clustering algorithm, outlined in Section \ref{sec:clustering}. To incorporate temporal information we multiply the textual affinity score between two tweets by a function exponentially decaying with the time gap.

\section{Clustering method}

\subsection{SMERC algorithm}
\label{sec:clustering}

\peter{The clustering algorithm which we present here aims to create clusters of tweets in response to the same event. We do this in a several-stage process, where most of the steps are standard text processing and clustering.} \caitlin{Our algorithm incorporates the novel use of temporal information in the form of an exponential decay function multiplied by the textual affinity between tweets.  This }\peter{prevents tweets that are a long time apart from being assigned to the same cluster.}
\caitlin{The clustering algorithm, including the text pre-processing steps, is summarized as follows and extended in detail below: }

\begin{enumerate}
\item Remove stop words from tweets.
\label{step:RemoveStopWords}
\item Convert words to their associated stems.
\label{step:StemWords}
\item Create a bag-of-words vector representing each tweet.
\label{step:BagOfWords}
\item Use TF-IDF to more heavily weight words that occur less frequently.
\label{step:TFIDF}
\item Use cosine similarity to determine the textual similarity between TF-IDF vectors representing tweets.
\label{step:CosineSimilarity}
\item Using the temporal information in each tweet, multiply by $e^{\Delta t_{ij} / T_p}$, an exponential decay function dependent on the time distance between tweets.
\label{step:ExponentialDecay}
\item Use affinity propagation to determine clusters.
\label{step:AffinityPropagation}
\item Filter clusters to ensure sufficiently high average affinity between elements.\\
\label{step:FilterClusters}
\end{enumerate}

\peter{We begin with $m$ tweets $\textbf{w} = \{w_i\}$ for $i = 1 ... m$. Steps \ref{step:RemoveStopWords} and \ref{step:StemWords}, the removal of stop words and taking the stem of each word are standard techniques used in natural language processing. }\caitlin{The pre-defined list of stop words in the Python Natural Language Toolkit (NLTK) \cite{Loper:2002:NNL:1118108.1118117} is used to remove words that are }\peter{purely for language structure, such as \emph{the} or \emph{at}.}

Stemming removes word suffixes such as \emph{ed} or \emph{ing} which do not provide much additional content meaning. 
This reduces the computational complexity by reducing the size of eventual word vectors, and allows easier identification of repeated word meanings. 
After the removal of stop words and stemming we are left with a reduced, cleaned, set of tweets ${\textbf{w}^{\prime}} = \{w_i^{\prime}\}$.\\

In Step \ref{step:BagOfWords}, we use bag-of-words to vectorize the tweets ${\textbf{w}^{\prime}}$ into a set of numerical vectors $\textbf{x} = \{x_i\}$ for $i = 1,\ldots,m$, where the elements of $x_i \in \mathbb{R}^n$ represent the count $f_{ji}$ for $j = 1,\dots,n$, of each word stem from a vocabulary $V$ of size $n$ that is used within tweet $w_i^{\prime}$.
We remark that this embedding of tweet $w_i^{\prime}$ is not essential; the techniques developed here can also be used with other vectorization methods such as word2vec \cite{DBLP:journals/corr/abs-1301-3781}. 
We used this standard method of vectorization for its simplicity and performance; experimentation showed that bag-of-words resulted in superior performance to word2vec when analyzing sporting events.
This is likely due to the context-specific word definitions used in sport-related tweets.
For example, we used the standard python implementation of word2vec\footnote{Available at https://github.com/danielfrg/word2vec.}, trained on Wikipedia data, and words such as \emph{tackle} likely have different meanings in sport than the Wikipedia contexts. 
If a unique dictionary was created for each sport, or if a sufficiently-large training corpus specific to the purpose could be created (eg., comprising sport-related tweets), it is very possible that word2vec would have superior performance.\\

%\begin{comment}
In Step \ref{step:TFIDF}, we use term frequency - inverse document frequency (TF-IDF):
\[
\text{TF-IDF} = \left({0.5 + 0.5\frac{f_{ji}}{\max_{k \in V}f_{ki}}}\right)\log \frac{m}{n_j}
\]
where $n_j$ is the number of documents in which word $j$ appears. This increases the weighting of words that are used infrequently in the document. For example, if an infrequently used word such as \emph{typhoon} occurs in only two tweets, it is likely that these tweets are related.\\
%\end{comment}

%In Step \ref{step:TFIDF}, we use term frequency - inverse document frequency (TF-IDF) to more heavily weight words that occur less frequently.
 
\peter{In Step \ref{step:CosineSimilarity}, we use cosine similarity to measure the similarity between any pair of tweets, giving a textual distance} 
\[
d(x_i,x_j) = d_{ij} =  \frac{x_i \cdot x_j}{||x_i|| ||x_j||}
\]
between two word vectors $x_i$ and $x_j$. Compared to other measures, such as Euclidean distance, this method emphasizes the impact of words that tweets have in common. 
%Cosine similarity removes the need for normalizing vectors as it gives us the angle between the two tweet vectors. 
We calculate the similarity between each pair of tweet vectors $x_i$ and $x_j$. We then have an $m \times m$ symmetric matrix $D$ of cosine similarities, where $d_{ij}$ is the cosine similarity between the $i$th and $j$th tweets.\\
	
As we showed from data in Section \ref{sec:ProbabilisticMotivation}, for two tweets separated by $\Delta t_{ij}$, the probability that the tweets are related is proportional to $e^{-\Delta t_{ij}/T_p}$ where $T_p$ is a constant. Consequently in Step \ref{step:ExponentialDecay}, for each pair of tweets $w_i$ and $w_j$ that occurred at times $t_i$ and $t_j$ respectively, we multiply the cosine similarity score by \caitlin{$e^{-\Delta t_{ij}/T_p}$} where $\Delta t_{ij} = |t_i - t_j|$ and $T_p$ is the time where the likelihood of tweets being related has fallen by the factor $1/e$. This gives us a matrix of exponential scaling factors $E$ where
\[
E_{ij} = e^{-\Delta t_{ij}/T_p}.
\]\\

In Step \ref{step:AffinityPropagation} we take the Hadamard (or element-wise) product of the matrix of cosine similarities $D$ with the exponentially decaying scaling factor matrix $E$, giving the resulting affinity matrix 
\[
C = D \circ E.
\]
%\begin{comment}
%\begin{equation}
%C=A \circ E.
%\label{eqn:C}
%\end{equation}
%\end{comment}
We use affinity propagation for the clustering itself, selecting the option for a pre-defined affinity matrix. Affinity propagation has order of complexity $O(k m^2)$ where $m$ is the number of elements, in our case tweets, and $k$ is the number of iterations.
 
The processes of calculating the cosine similarity between each pair of tweets, mulitplying by an exponential decay function, and undertaking the clustering by affinity propagation are all operations with time complexity $O(m^2)$ where $m$ is the number of input tweets. These will potentially slow processing for larger tweet collections. However, as the calculations for each pair of tweets is independent, this process can be sped up by parallelizing the algorithm and evaluating the exponential function on GPUs instead of CPUs. If the number of tweets in the dataset is still too large for efficient processing, we can split the input dataset into smaller windows of time. As our tweet clustering algorithm avoids putting tweets with high temporal distance in the same cluster, such an action will only have an effect on the output around the time window boundaries.\\

In Step \ref{step:FilterClusters}, after the clustering is complete, we measure the internal average affinity between elements. If this is above a threshold value $\delta$, we keep the cluster. Testing for this threshold is necessary for the purpose of event prediction as the affinity propagation algorithm assigns all tweets to a cluster. Consequently, there will be a number of clusters containing tweets that are unrelated to specific events.
While these clusters are informative about the background topics of conversation taking place during a sporting contest, for the purpose of event prediction they are discarded. Our method filters out any such clusters.\\

\peter{Previous clustering methods are inferior at dealing with temporal information. Generally time is used as a linear variable which incorporates the loss of interest in topics over time }\caitlin{less accurately.  Another benefit of} \peter{our} \caitlin{method is that} \peter{ affinity propagation does not require selection of the number of clusters beforehand. It is entirely unsupervised and generates clusters of tweets related to events without requiring human assistance.}

\section{Experiments}

%\peter{We test our proposed tweet clustering method and compare it to existing methods.}

We now test our proposed tweet clustering method and compare it to the following exisiting methods: Twevent \cite{Li:2012:TSE:2396761.2396785}, Twitinfo \cite{marcus2011twitinfo}, and MABED \cite{DBLP:journals/corr/GuilleF15}.
We run our clustering algorithm SMERC on our three collected datasets. Through manually examining the cluster output, we determine whether each individual cluster is related to a real event. We then calculate the precision by dividing the number of clusters about real events, by the total number of clusters. An event is considered \emph{missed} if our clustering algorithm conceivably could have detected the event, but failed to do so. We calculate the recall in the standard way, by dividing the number of true detected events by the total number of true events.

Due to the unavailability of the datasets used to test exisiting methods as well as usable code for these methods, we can neither test our method on previous authors' datasets, nor apply their methods to the datasets collected here. We will therefore reproduce the authors' precision and recall reported in their papers, and compare with the same values we calculate here. These values should therefore only be taken as indicative of the overall skill of each method. We will make the datasets used here, as well as code for SMERC, available upon acceptance of this paper.

\subsection{Results}

\begin{table}%[h]
	\caption{Comparison of event detection precision and recall between our method and others. Note that as we are testing on different datasets, these results are indicative only.}
	\begin{center}
		\begin{tabular}{l | c | c | c | c}
			{Method} & {Tweet topic} & {Precision} & {Recall}  & {F1 score} \\
			\hline
			SMERC & AFL and cricket & \textbf{0.866} & 0.724 & \textbf{0.789} \\
			Twitinfo \cite{marcus2011twitinfo} & Soccer events & 0.773 & \textbf{0.773} & 0.773 \\
			Twevent \cite{Li:2012:TSE:2396761.2396785} &  (Singaporean) news & 0.762 & 0.619 & 0.683 \\
			MABED \cite{DBLP:journals/corr/GuilleF15} & French politics &  0.775 & 0.609 & 0.682 \\
%			Ununkard \emph{et al.} (LSED) \cite{Unankard2015} & Queensland and earthquakes & \textbf{0.864 to 0.973} & 0.014 to 0.662 \\			
		\end{tabular}
	\end{center}
	\label{tab:PrecisionRecallSummary}
\end{table}

\begin{table}%[h]
	\caption{Improvement to cluster quality (the proportion of tweets within a cluster related to the event of interest) with temporal adjustment. This significant improvement is the key benefit provided by our method.}
	\begin{center}
		\begin{tabular}{l | p{14mm} | p{15mm}}
			{Dataset} & {No temporal adjustment} & {With temporal adjustment} \\
			\hline
			D1: AFL first prelim final & \centering 0.455 & \textbf{0.887}\\
			D2: Heat vs Stars BBL game & \centering 0.567 &  \textbf{0.921} \\
			D3: WBBL first weekend &  \centering 0.568 &  \textbf{0.971} \\
		\end{tabular}
	\end{center}
	\label{tab:QualitySummary}
\end{table}

In Table \ref{tab:PrecisionRecallSummary}, we compare the precision and recall of our method to other event detection techniques. As standard test sets are not available, each method is evaluated on its own datasets.
Despite this, and even though our method SMERC is tested on an arguably more challenging dataset where ``events'' are difficult to define, 
it performs remarkably well,
 showing the highest F1 score of all methods.
 SMERC also shows the highest precision of all methods, and second-highest precision. 
%However, despite using more challenging datasets, our method SMERC recorded state-of-the-art precision and recall scores.
We use the Twevent \cite{Li:2012:TSE:2396761.2396785} approach for evaluating precision and recall, comparing the output clusters to real events. 
 Of course, we note that as we are comparing on different categories of datasets, these results are indicative only. 
 A better evaluation of our method is to investigate the quality of individual clusters,
 by looking at the types of tweets that are grouped together, particularly with the temporal adjustment.

However, our purpose in this paper is not merely a method that achieves high precision and recall statistics, but one which generates meaningful clusters about events. We therefore define the \emph{quality} of a cluster as the proportion of tweets within a cluster which are related to the topic of the overall cluster. The improvement in the quality of the clustering from our temporal adjustment is determined by comparing average cluster quality both with and without the temporal adjustment. 
%Furthermore, we will analyse the content of, and present example tweet clusters, to show how our method selects individual tweets for inclusion or exclusion from clusters.

Table \ref{tab:QualitySummary} shows the increase in cluster quality after our temporal adjustment. 
This demonstrates the importance of the temporal adjustment as characterized by the exponentially-decaying model of user interest employed here.
When applying this transformation to the affinity matrix $A$, the cluster quality increases substantially from around 50-60\% to around 90-100\% across all datasets. 
This is the primary benefit provided by our method: temporally distant tweets from unrelated events tend to not be mixed with clusters of tweets about a single event, and incorporating this exponential decay in interest improves the results dramatically.
We now explore in greater detail the nature of the tweet clusters found by our method.

\subsection{Example clusters}

We give two examples demonstrating the effectiveness of our method on our collected Twitter datasets. We firstly examine the output of our method on tweets about Australian Rules Football, for the game between the Adelaide Crows and the Geelong Cats on 22 September 2017, dataset D1. We ran the clustering algorithm both with (Table \ref{tab:AFLClusteredTweetsWith}), and without (Table \ref{tab:AFLClusteredTweetsWithout}), the temporal adjustment. As can be seen, both methods correctly clustered a series of tweets following a goal at around 09:54 GMT by the Adelaide Crows Football Club player Eddie Betts. However, without the temporal adjustment, tweets about an additional goal at around 10:14 GMT were also clustered together. It is clear by watching the games and looking at the tweet times that these were separate events.
When using the temporal adjustment encapsulated by the exponentially-decaying scaling factor matrix $E$, our method correctly separates the tweets into appropriate groupings.

\begin{table*}
		\caption{AFL dataset example cluster created using our method SMERC. These tweets all relate to an event occuring at 09:54 GMT.}
	\begin{center}
		\begin{tabular}{l | c}
			Tweet body & Time (GMT), 22 Sep 2017\\
			\hline
			Eddie!!!! \#AFLCrowsCats & 09:54:11 \\
			Eddie, you beauty!!! \#AFLCrowsCats & 09:54:17 \\
			\#AFLCrowsCats sorry cats fans... I LOVE EDDIE! & 09:54:20 \\
			Fair shark by Eddie. \#AFLCrowsCats & 09:54:32 \\
			\#AFLCrowsCats Eddie's Best & 09:54:51 \\
			Eddie's goal from a stoppage was a coach killer! Can't let him move like that in F50! \#AFLCrowsCats & 09:56:02
		\end{tabular}
		\label{tab:AFLClusteredTweetsWith}
	\end{center}
\end{table*}

\begin{table*}
		\caption{Tweets that were removed from the cluster in Table \ref{tab:AFLClusteredTweetsWith} by our method due to the temporal difference. These tweets all relate to a similar goal event by Eddie Betts, but occuring 10:14 GMT.}
	\begin{center}
		\begin{tabular}{l | c}
			Tweet body & Time (GMT), 22 Sep 2017\\
			\hline
			Eddie! What a goal! 37-8  \#AFLCrowsCats & 10:14:34 \\
			EDDIE.************.BETTS.\#AFLCrowsCats & 10:14:36 \\
			It's Eddie's world and we're just living in it \#AFLCrowsCats & 10:14:37 \\
			Eddie! You are the king of Adelaide! \#AFLCrowsCats & 10:14:38 \\
			Uncle Eddie, ******* hell. \#AFLCrowsCats & 10:14:39 \\
			Eddie. What more can you say? \#AFLCrowsCats & 10:14:45 \\
			That was delicious, Eddie! \#AFLCrowsCats & 10:14:45 \\
			Eddie. Betts. He is that good! \#AFLCrowsCats & 10:14:47 \\
			Eddie. \#AFLCrowsCats \#WeFlyAsOne & 10:14:48 \\
			\#AFLCrowsCats Eddie's on fire & 10:14:54 \\
			Beautiful Eddie. Beautiful. @Adelaide\_FC \#AFLCrowsCats \#AFLFinals & 10:15:07
		\end{tabular}
		\label{tab:AFLClusteredTweetsWithout}
	\end{center}
\end{table*}

A second example of the output of our tweet clustering method is also given. Table \ref{tab:WBBLClusteredTweetsWith} gives six tweets about when Perth Thunder cricketer Rachael Haynes scored 50 runs, a significant milestone in this form of the game and notable event in this particular match. Table \ref{tab:WBBLClusteredTweetsWithout} gives the three tweets that were removed by our method due to the temporal difference.
We note that all three tweets were unrelated to the event of Haynes' 50-run milestone: the first is of an earlier 6-run shot by Haynes, and the other two are for 50-run milestones scored by other players over an hour from the event of interest. These are correctly separated out by the temporal adjustment.

\begin{table*}
		\caption{WBBL dataset example cluster created using our method SMERC}
	\begin{center}
		\begin{tabular}{l | c}
			Tweet body & Time (GMT), 9 Dec 2017\\
			\hline
			50 for Haynes and it comes from just 37 balls 4/104 \#ThunderNation \#WBBL03 & 14:10:50 \\
			What a knock from Haynes, she reaches 50 off just 37 balls for @ThunderWBBL \#WBBL03 & 14:11:34 \\
			50 for @RachaelHaynes in just 37 balls. The first fifty of \#WBBL03 & 14:12:02 \\
			\#WBBL03 First FIFTY of the season. The Aussie skipper @RachaelHaynes brings it up in 37 balls. & 14:13:55 \\
			\#WBBL03 First FIFTY of the season. @RachaelHaynes brings it up in 37 balls. & 14:15:32 \\
			\#WBBL03 First FIFTY of the season. @RachaelHaynes brings it up in 37 balls. & 14:17:40
		\end{tabular}
		\label{tab:WBBLClusteredTweetsWith}
	\end{center}
\end{table*}

\begin{table*}
		\caption{Tweets that were removed from the cluster in Table \ref{tab:WBBLClusteredTweetsWith} due to the temporal difference. Note that all tweets are related to different events.}
	\begin{center}
		\begin{tabular}{l | c}
			Tweet body & Time (GMT), 9 Dec 2017\\
			\hline
			That's what happens when my favourite leftie @RachaelHaynes  Middles the ball..  First six in \#WBBL03 & 13:57:01 \\
			A maiden @WBBL fifty for @Jess\_cameron27. This is also the second fifty of this season! & 15:50:33 \\
			50 in just 22 balls. That's @ashleighgardne2 for you. \#WBBL03 &  18:12:27
		\end{tabular}
		\label{tab:WBBLClusteredTweetsWithout}
	\end{center}
\end{table*}

As can be seen in Table \ref{tab:DatasetSummary}, the filtering of clusters to remove spam tweets and to ensure a sufficient average pairwise tweet affinity removes over half of the clusters. In particular, for the WBBL dataset which was collected over a weekend, time gaps between tweets are often higher. Consequently, a very high number of clusters are removed by the filtering.

The tradeoff between precision and recall can be controlled through the minimum average pairwise affinity between tweets. Setting a higher value for this parameter increases precision but reduces recall, and vice-versa.

\begin{table*}
	\caption{Dataset clustering summary. Filtering refers to the process of removing bot tweets and clusters with an insufficently high average pairwise affinity. The WBBL dataset has a higher percentage of clusters removed by this mechanism as it was collected over a longer time period.}
	\begin{center}
		\begin{tabular}{l | p{16mm} | p{16mm} | p{16mm} | p{16mm} | p{16mm} | p{16mm} | p{16mm} }
			\textbf{Dataset} & \textbf{\# Collected tweets} & \textbf{\# Clusters before filtering} & \textbf{\# Clusters after filtering}  & \textbf{\# Clusters linked to specific events} & \textbf{\# Missed events} & \textbf{Precision} & \textbf{Recall} \\
			\hline
			D1: AFL first prelim final & \centering 5393 & \centering 277 & \centering 152 & \centering 128 & \centering 50 & \centering 0.842 & \hspace{2mm} 0.719 \\
			D2: Heat vs Stars BBL game & \centering 5018 & \centering 430 & \centering 160 & \centering 144 & \centering 45 & \centering 0.900 & \hspace{2mm} 0.762 \\
			D3: WBBL first weekend & \centering 3153 & \centering 260 & \centering 42 & \centering 36 & \centering 16 & \centering 0.857 & \hspace{2mm} 0.692 \\
		\end{tabular}
	\end{center}
	\label{tab:DatasetSummary}
\end{table*}

\section{Conclusions}

In this paper we have developed a new method, \emph{Social Media Event Response Clustering} (SMERC), for clustering tweets by using temporal information. This exponentially-decaying model was informed by a detailed analysis of a number of Twitter datasets collected around sporting events, and is critical for improving the quality of clusters generated by the method.

In addition to sport, SMERC could be applied to other social science fields such as social unrest or natural disasters, where people respond to real world events. Also, in addition to Twitter data, this work could be applied to data from other social media platforms such as Facebook or Youtube. Future work will test SMERC to such data streams, as they become available. 

At a computational level, there is also scope for future work in reducing the time complexity of the algorithm. SMERC could also be adapted into a system to detect events in real time, rather than using post processing. Furthermore, creating a dedicated dictionary for the topics that we are analysing would potentially allow the use of more modern tweet processing methods such as Tweet2Vec \cite{P16-2044}.

\peter{The difficulty of event detection varies depending on the topic of the event. Detecting rare events with known keywords such as earthquakes or goals in soccer, is much easier than detecting less well defined or frequently occurring events. After an earthquake, many people will tweet the word ``earthquake'', a word which is rarely mentioned otherwise. Consequently, the choice of dataset will affect the measured performance of the algorithms.}

Our clustering and time estimation approaches could potentially by improved by automated removal of noise on the input Twitter data \cite{Nasim2018}, as done for other purposes \cite{Liu:2016:RTL:2983323.2983363}. Spam and advertisements tend to repeat identical or very similar tweets, which have high pairwise text similarity. Automated removal of these tweets before clustering will improve the amount of information content in the output clusters.

\peter{Unfortunately, obtaining code or datasets for tweet clustering and event detection methods is often prohibitive, which limits the ability to directly compare new methods to the existing state-of-the-art. This in turn makes determining what in fact constitutes ``state-of-the-art'' for a particular application effectively impossible. In the interests of reproducibility of work and having common tweet sets for testing, we will make our datasets and code public and encourage other authors to do the same.}

From a theoretical perspective, our work improves the understanding of the distribution of decay of interest reflected in online social media data streams around events. Practically, it provides an effective method to cluster tweets for the purpose of event detection. We evaluated this both quantitatively through the calculation of standard evaluation metrics such as precision and recall, but also qualitately through inspection of the actual tweets clustered together by our method. We believe our method could be deployed by governments or other entities to conduct social sensing from microblogs.

%\clearpage
\bibliographystyle{IEEEtran}
\bibliography{bibliography}

\end{document}